\documentclass{iaus}
\usepackage{graphicx}

\title[Magnetic fields in classical Be stars]
{Magnetic fields in classical Be stars}

\author[R. Yudin, S. Hubrig, M. Pogodin et al.]
{R.~Yudin$^1$,
 S.~Hubrig$^2$, M.~Pogodin$^1$, I.~Savanov$^3$, M.~Sch\"oller$^4$, G.~Peters$^5$ \and M.~Cure$^6$}

\affiliation{$^1$Pulkovo Observatory, St.-Petersburg, Russia; \break
email: ruslan61@hotmail.com; $^2$ESO, Santiago, Chile; $^3$Institute
of Astronomy, Russian Academy of Science, Moscow, Russia $^4$ESO,
Garching, Germany; $^5$Univ. Southern California, USA; $^6$Univ. de
Valparaiso, Chile }

\pubyear{2008}
\volume{259}  
\pagerange{1--2}
\date{?? and in revised form ??}
 \jname{Cosmic Magnetic Fields: from Planets,
to Stars and Galaxies } \editors{K.G. Strassmeier, A.G. Kosovichev
\& J. Beckmann, eds.}
\begin{document}

\maketitle

\begin{abstract}
We report the results of our study of magnetic fields in a
sample of Be stars using spectropolarimetric data
obtained at the European Southern Observatory with the multi-mode
instrument FORS\,1 installed at the 8\,m Kueyen telescope. The detected
magnetic fields are rather weak, not stronger than
$\sim$150\,G. A few classical Be stars display
cyclic variability of the magnetic field with periods
of tens of minutes.

\end{abstract}

\section{Introduction}
Over an extended period of years many attempts to obtain reliable
direct measurements of magnetic fields of Be stars have been rather
unsuccessful. Recently, weak field detections in a few Be stars were
reported by \cite{Hubrig07}. Below, we report some preliminary
results of our most recent observing runs with the multi-mode
instrument FORS1 installed at the 8m Kueyen telescope at the VLT. We
also present first results of our magnetic field time series in the
Be star $\lambda$ Eri.

\section{Observations}\label{sec:observ}
\begin{figure}
\centering
\includegraphics[height=1.8in,width=2.6in,angle=0]{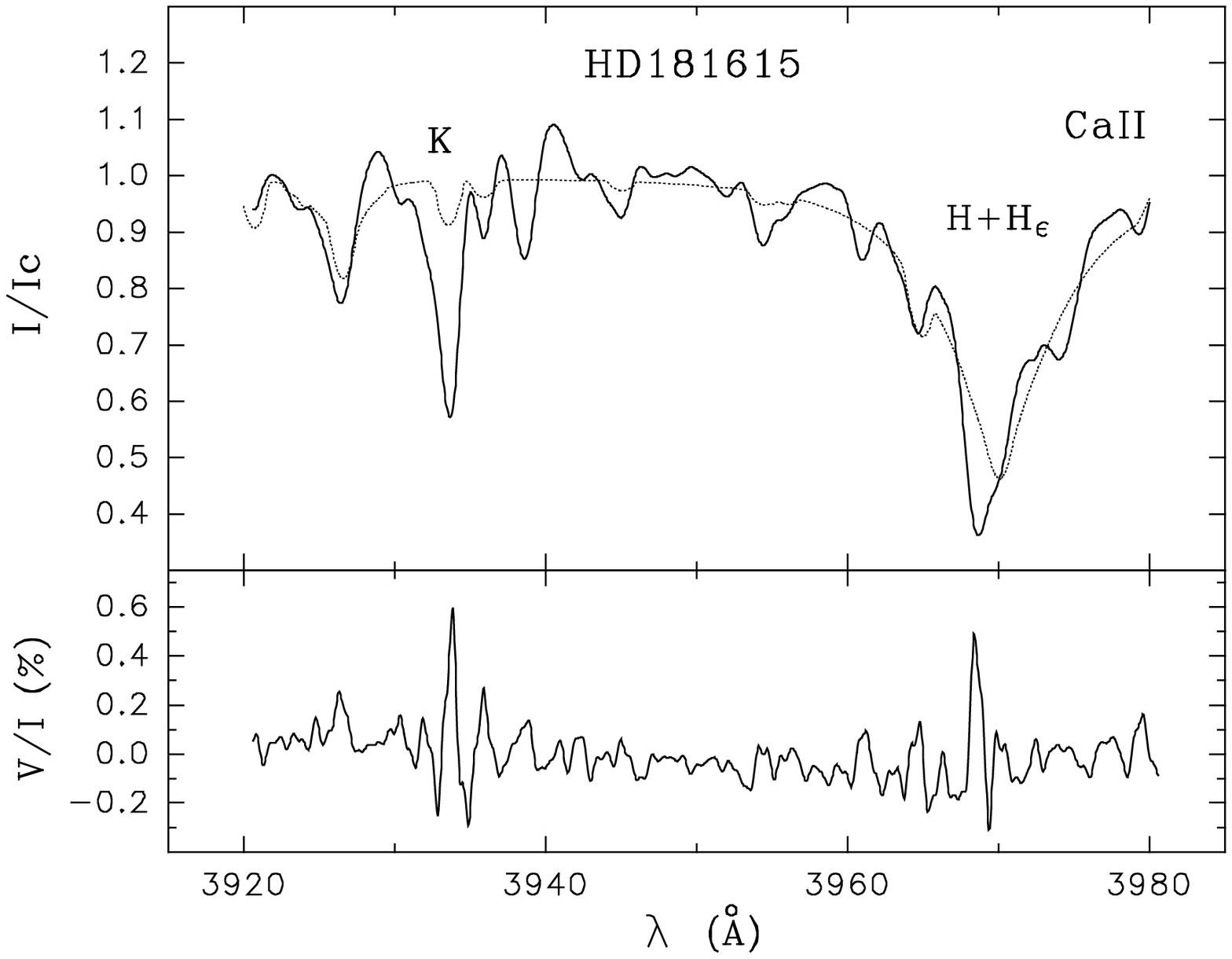}
  \includegraphics[height=1.8in,width=2.6in,angle=0]{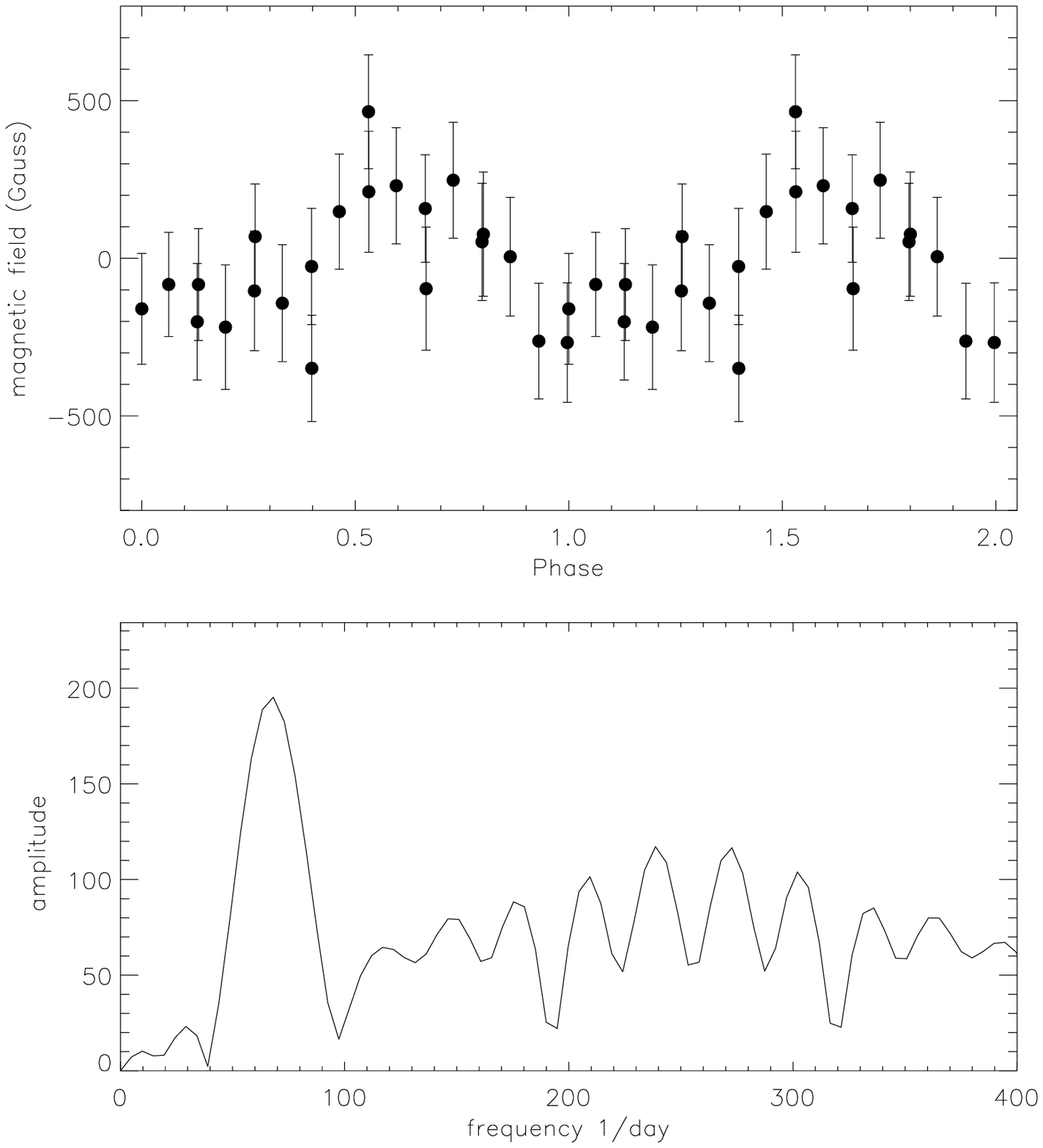}
 \caption{Left panel: Stokes~$I$ and $V$ spectra in HD\,181615;
Right panels: Cyclic variability of the magnetic field in $\lambda$
Eri with a period of 21.1\,min and Fourier spectrum for the magnetic
field data derived from hydrogen lines.
 } \label{fig:Ca}
\end{figure}

As it was already reported by \cite{Hubrig07}, weak photospheric
magnetic fields have been detected in four stars, HD\,56014,
HD\,148184, HD\,155806, and HD\,181615. We also pointed out the
likely presence of distinct circular polarisation features in the
circumstellar components of Ca II H\&K in the three stars,
HD\,58011, HD\,117357, and HD\,181615, indicating a probable
presence of magnetic fields in the circumstellar mass loss disks.
The FORS\,1 observations of HD\,181615 are presented in the left
panel of Fig.~\ref{fig:Ca}. New observations of classical  Be stars
have been been carried out at the VLT from 2006 to 2008. The Be
stars have been observed in 2006 in service mode with the GRISM 600B
in the wavelength range 348-589\,nm and with the GRISM 1200B in the
spectral range 373-497\,nm. The last observations we report here
have been carried out in visitor mode in 2008 in the framework of
the study of 15 B-type members of the open cluster NGC~3766.

\section{Results}
 Weak magnetic fields have been detected in eight classical
Be stars. The magnetic field strengths of the studied Be stars are
found to be in the range 40\,G to 150\,G. A few Be stars with
detected magnetic fields show non-detections at some other observing
dates, indicating possible temporal variability of their magnetic
fields. To confirm this suggestion we performed time-resolved
magnetic field measurements over one hour for a few Be stars to get
information about the behaviour of the magnetic field over at least
a part of the stellar surface. The service mode observations
revealed periodic changes of magnetic fields on the time scale of
tens of minutes in a few classical Be stars. Among these stars, the
star $\lambda$ Eri displays a cyclic variability of the magnetic
field with a period of 21.1\,min (see the right panel of
Fig.~\ref{fig:Ca}). The results of the study of fifteen early B-type
members of the open cluster NGC~3766 have already been reported by
\cite{McSwain08}  who noted two definite and two possible detections
in the studied sample. We detected three additional stars with weak
magnetic fields which are members of this cluster (highlighted by
bold face in Table~\ref{tab:NGC}).

\section{Conclusions}\label{sec:concl}
Our search for magnetic fields in Be stars revealed that their
 magnetic fields are rather weak, but fields of less than $\sim$150\,G
are not rare.
 The magnetic fields are clearly variable and a non-detection of magnetic fields in some stars
 may be explained by temporal variability.
 A few classical Be stars display cyclic variability of the magnetic field with periods
 of tens of minutes.
 The cluster NGC~3766 seems to be extremely interesting, where we find clear evidence for the
 presence of a magnetic field in seven early B-type cluster members out of fifteen members.

\begin{table}\def~{}
  \begin{center}
  \caption{Measurements of NGC 3766 cluster stars with detected magnetic fields.}
  \label{tab:NGC}
  \begin{tabular}{cccc}\hline
      NGC~3766  & MJD   & $\left< B_z\right>_{\rm all}$ & $\left< B_z\right>_{\rm hydr}$ \\
 \hline

45  & 54550.066  &  $-$78$\pm$41 & $-$194$\pm$62\\
47 &54549.020 &$-$146$\pm$43& $-$129$\pm$58\\
94 & 54550.327  & 294$\pm$53& 310$\pm$65 \\
{\bf 111} &  54549.020  & {\bf 112$\pm$34}& 89$\pm$38 \\
170 & 54550.186 & 1522$\pm$34&  1559$\pm$38\\
{\bf 176} & 54550.016& {\bf 121$\pm$36}& {\bf 141$\pm$41}\\
{\bf 200} & 54550.375 &  {\bf 128$\pm$40} & {\bf 115$\pm$38}\\
\hline
  \end{tabular}
 \end{center}
\end{table}

\end{document}